%% file: juettner_hadron2011.tex
\documentclass[a4paper,11pt]{article}

\usepackage{contribution}
\usepackage{epsfig,psfrag,multirow,color,rotate,rotating}

\input{econfmacros}
\input{contributionmacros}

\begin{document}

\input{contribution}

\end{document}

%% file: econfmacros.tex
\newcommand{\weblink}[2][]{%
    \ifthenelse{\equal{#1}{}}%
    {\textnormal{\url{#2}}}%
    {\textnormal{\href{#2}{#1}}}%
}

\newcommand{\acknowledgements}[1]{%
  \bigskip\bigskip
  \textsf{\textbf{\Large Acknowledgements}} \\[2ex]
  {#1}
  \bigskip
}

\def\beq{\begin{equation}}
\def\eeq#1{\label{#1}\end{equation}}
\def\eeqn{\end{equation}}

\def\beqa{\begin{eqnarray}}
\def\eeqa#1{\label{#1}\end{eqnarray}}
\def\eeqan{\end{eqnarray}}

\let\bar=\overbar

\def\Dslash{\not{\hbox{\kern-4pt $D$}}}
\def\dslash{\not{\hbox{\kern-2pt $\del$}}}

\def\msb{{\bar{\ssstyle M \kern -1pt S}}}

%% file: contributionmacros.tex
\newcommand{\contribution}[7][]{%
  \clearpage
  \thispagestyle{plain}
  \ifthenelse{\equal{#1}{}}
  {\hypersetup{pdftitle={#2}}}
  {\hypersetup{pdftitle={#1}}}
  \hypersetup{pdfauthor={{#3} {#4}}}
  {\centering\normalfont\LARGE\bfseries\sffamily #2 \par\nobreak}
  \lhead{}
  \chead{%
    \textit{\footnotesize XIV International Conference on Hadron Spectroscopy
      (\weblink[\textit{hadron2011}]{http://www.hadron2011.de}), 13-17 June 2011, Munich, Germany}%
  }
  \rhead{}
  \bigskip
  \begin{center}
    {#3} {#4}\ifthenelse{\equal{#6}{}}{}{\footnote{\weblink[#6]{mailto:#6}}}
    \ifthenelse{\equal{#7}{}}{}{#7} \\
    \textit{#5}
  \end{center}
  \bigskip
}

\renewcommand{\abstract}[1]{%
  \begin{center}
    \begin{minipage}{0.85\textwidth}
      \begin{footnotesize}
        #1
      \end{footnotesize}
    \end{minipage}
  \end{center}
  \bigskip
}

%% file: contribution.tex
%
%
%
%
%
{  

\def\bi{\begin{itemize}}
\def\ei{\end{itemize}}
\def\tbr{\hspace{0.25mm}{\color{red}$\blacksquare$}} 
\def\good{{\color{green}$\bigstar$}}
\def\tbg{{\color{green}$\bigstar$}}
\def\soso{\hspace{0.25mm}{\color{orange}\Large\textbullet}}
\def\tbo{\hspace{0.25mm}{\color{orange}\Large\textbullet}}
\def\bad{\raisebox{0.35mm}{\hspace{0.65mm}{\color{red}\tiny$\blacksquare$}}} 
\def\rC{{\color{black}\small \tt C}}
\def\gA{{\color{black}\small \tt A}}
\def\oP{{\color{black}\small \tt P}}
\def\Vus{V_{us}}
\def\fplus{f_+(0)}
\definecolor{orange}{rgb}{1.0,.6,0}

%

\contribution[Review: The FLAG working group]  
{Review: The FLAG working group}
{Andreas}{J{\"u}ttner}  
{speakers affiliation:\\CERN, Physics Department, TH Unit\\
CH-1211 Geneva 23, Switzerland}  
{juettner@mail.cern.ch}  
{on behalf of the FLAG Collaboration}  
%
\abstract{
The FLAG working group
reviews lattice results relevant for pion and kaon physics with the aim
of making them easily accessible to the particle physics phenomenology community.
The set of quantities considered so far comprises 
light quark masses,
kaon and pion form factors, the kaon mixing parameter, and low energy constants
of SU(2)$_L\times$SU(2)$_R$ and SU(3)$_L\times$SU(3)$_R$ chiral perturbation theory.
}
\vskip -10.5cm 
\begin{minipage}{\linewidth}
\begin{flushright}
CERN-PH-TH/2011-207
\end{flushright}
\end{minipage}
\vskip 10cm
%
\section{Introduction}
Strong claims like\\[-8mm]
\bi
 \item ``We find a (2-3)$\sigma$ tension in the unitarity triangle''~\cite{Laiho:2009eu} \\[-8mm]
 \item ``\dots~confirming CKM  unitarity at the permille level''~\cite{Colangelo:2010et}\\[-8mm] 
 \item ``\dots~we find evidence of new physics in both $B_d$ and $B_s$ systems''~\cite{Lenz:2010gu} \\[-8mm]
 \item ``Possible evidence for the breakdown of the CKM-paradigm of \linebreak CP-violation''~\cite{Lunghi:2010gv}\,,
\ei\mbox{}\\[-8mm]
 have recently been made based on lattice QCD results.
This 
illustrates the role  lattice QCD predictions are playing for the phenomenology
of the Standard Model (SM). To this end the FLAG 
working group has
formed with the aim of allowing also to an outsider a judgement of the
quality and \textit{state-of-the-art-fulness} of lattice results.

The quantities FLAG is currently considering  comprise the light quark masses
$m_u$, $m_d$, $m_s$, the ratio of the leptonic kaon and pion decay constant,
$f_K/f_\pi$, the semi-leptonic $K\to\pi$ form factor at vanishing momentum
transfer $\fplus$, the kaon mixing parameter $B_K$ and 
a number of low energy constants of 
SU(2)$_L\times$SU(2)$_R$ and SU(3)$_L\times$SU(3)$_R$ chiral perturbation theory.

The FLAG report provides all relevant formulae and notation, a detailed quality
assessment of every computation, where FLAG considers appropriate an
average or recommended range and a lattice dictionary for non-experts and details
of every single lattice simulation. After having made a start FLAG is
currently discussing extending the set of quantities (e.g. charm and beauty).  
Periodic updates of the results are planned. Note also the similar
effort by \cite{Laiho:2009eu}.

\section{Lattice QCD}\mbox{}\\[-7mm]
While perturbation theory is an invaluable tool at weak coupling it cannot
make predictions for bound state observables like the proton mass or the properties
of hadron decays. While potential models and sum rules clearly allow for 
studying these properties, only lattice QCD has the potential to make
systematically improvable predictions.

QCD is a theory with three colour charges, six quarks in the fundamental 
representation with broken iso-spin in an infinite space-time continuum with
the charged pion mass being 139.6MeV. In
order to regularise the theory and make it accessible to simulations by
means of a Monte-Carlo integration of its defining path integral, in lattice
QCD one reduces its dynamical 
flavour content to $N_f=2$ (degenerate up- and down-quark),
2+1 (plus a strange quark) or 2+1+1 (plus a charm quark), keeps the iso-spin
symmetry exact, varies the pion mass ideally close to the physical point, 
considers only a finite volume of around 3-4fm and discretises space time 
with a lattice spacing of typically down to $0.05$fm.

Clearly lattice QCD is thus affected by a number of systematic effects which
need to be understood theoretically  and controlled in the numerical simulation.
For each of the major systematic effects FLAG has laid our quality criteria
which are explained in detail in the document. In order to allow the reader
a rough screening of available results, summaries are provided in terms of a
colour coding:\\[-8mm]
\bi
\item[\good] when the systematic error has been estimated in a 
satisfactory manner and convincingly shown to be under control;
\item[\soso] when a reasonable attempt at estimating the systematic 
error has been made, although this could be improved;
\item[\bad] when no or a clearly unsatisfactory attempt at 
estimating the systematic error has been made.
\ei\mbox{}\\[-8mm]
This colour coding has been set up for judging the quality
of the chiral extrapolation, the continuum extrapolation, finite size effects,
renormalisation, renormalisation scale running and the publication status.
Where applicable, results only enter averages if they carry no red tag for 
any of the criteria. This also applies to the publication status - only 
published and peer-reviewed results (or simple updates of previously published data and
analyses)
qualify for inclusion into any average.
FLAG treats results with different flavour content $N_f$ separately.\\[-8mm]
\section{FLAG summaries}\mbox{}\\[-7mm]
In this section a selection of results of the FLAG document with a focus 
on the determination of the CKM-matrix element $V_{us}$ are 
highlighted. Note
that the summaries are based on the FLAG document~\cite{Colangelo:2010et}. 
Results that 
have appeared in the meantime will be included in future updates of the
document.
\subsection{CKM first-row unitarity - $|V_{us}|$}\mbox{}\\[-7mm]
The determination of $|\Vus|$ proceeds as follows: On the one hand,
one experimentally
measures the rate of  a flavour changing process 
$s\to u$, where $s$ is the strange quark
and $u$ the up quark. On the other hand one computes the SM
prediction for the same process whose amplitude is proportional to the
CKM-matrix element $|\Vus|$ and which receives con\-tribu\-tions  from  the
electromagnetic, the weak and the strong interactions. While the former two 
are treated in perturbation theory for the processes
considered here, the contribution from the latter 
 needs to be computed in lattice QCD.
Eventually, $|\Vus|$ is  determined by equating the experimental result with
the SM-prediction. \\
For kaon/pion leptonic decays the relation between experiment and theory
in the SM was computed by Marciano~\cite{Marciano:2004uf} and using the
latest analysis of experimental results~\cite{Antonelli:2010yf} 
it yields the correlation
  	\begin{equation}\label{eqn:masterfKpi}
		\left|\frac{\Vus f_K}{V_{ud}f_\pi}\right|=0.2758(5)\,.
  	   \end{equation}
Since lattice QCD can provide $f_K/f_\pi$ one obtains a prediction for
$|\Vus/V_{ud}|$.
For semi-leptonic kaon decays the latest summary of experimental results 
together with SM-con\-tribu\-tions yields~\cite{Antonelli:2010yf}
\begin{equation}\label{eqn:masterKtopi}
 |\Vus| \fplus=0.2163(5) \,,
\end{equation}
and $|\Vus|$ is readily extracted provided a prediction of $\fplus$.\\
Figure \ref{fig:scatter} exemplifies FLAG's compilation of results showing the
 lattice data with 2, 2+1 and 2+1+1 flavours of dynamical 
fermions for both $\fplus$ and $f_K/f_\pi$. As an example we show the underlying
data for $\fplus$ together with the FLAG assessment in table \ref{tab:scatter}.
\begin{figure}
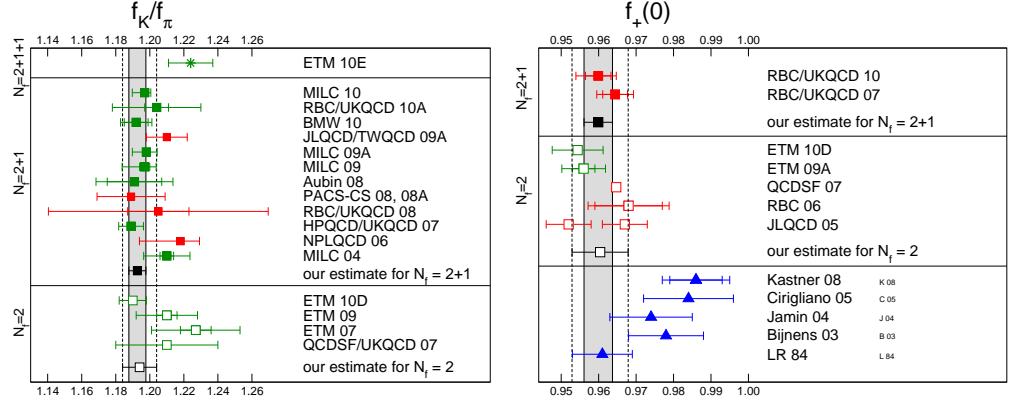

\begin{center}
\psfrag{x}[lc][l][1][0]{}
\psfrag{y}[lc][l][1][0]{}
\epsfig{scale=.35,file=bildla/RKpi.eps}\hspace{2mm}
\epsfig{scale=.35,file=bildla/f+0.eps}
\end{center}
\caption{Scatter plots  
of the lattice data on the ratio of leptonic
decay constants $f_K/f_\pi$ and the semi-leptonic $K\to\pi$
vector form factor at vanishing momentum transfer, $\fplus$. Green symbols
identify results that are free of \textit{red tags} according to FLAG's
assessment. The vertical bands correspond to FLAG's average/recommended range
for $N_f=2$ and $N_f=2+1$, respectively.
	\label{fig:scatter}}
\end{figure}
\begin{table}

\mbox{}\\[1.9cm]
\footnotesize
{\begin{tabular*}{\textwidth}[l]{@{\extracolsep{\fill}}llllllll}
Collaboration & $N_f$ & 
\hspace{0.15cm}\begin{rotate}{60}{publication status}\end{rotate}\hspace{-0.15cm}&
\hspace{0.15cm}\begin{rotate}{60}{chiral extrapolation}\end{rotate}\hspace{-0.15cm}&
\hspace{0.15cm}\begin{rotate}{60}{continuum extrapolation}\end{rotate}\hspace{-0.15cm}&
\hspace{0.15cm}\begin{rotate}{60}{finite volume errors}\end{rotate}\hspace{-0.15cm}&\rule{0.3cm}{0cm}
$f_+(0)$ \\
&&&&&&& \\[-0.1cm]
\hline
\hline&&&&&&& \\[-0.1cm]
RBC/UKQCD 10              &2+1  &\gA&\soso&\tbr&\tbg& 0.9599(34)($^{+31}_{-47}$)(14)\rule{0cm}{0.4cm}\\ 
RBC/UKQCD 07              &2+1  &\gA&\soso&\tbr&\tbg& 0.9644(33)(34)(14)\\
&&&&&&& \\[-0.1cm]
\hline
&&&&&&& \\[-0.1cm]
ETM 10D                    &2 &\rC&\soso&\tbg&\soso& 0.9544(68)$_{stat}$\\
ETM 09A 	           &2 &\gA&\soso&\soso&\soso& 0.9560(57)(62)\\	
QCDSF 07	           &2 &\rC&\tbr&\tbr&\tbg& 0.9647(15)$_{stat}$ \\
RBC 06  	           &2 &\gA&\tbr&\tbr&\tbg& 0.968(9)(6)\\	
JLQCD 05 	           &2 &\rC&\tbr&\tbr&\tbg& 0.967(6), 0.952(6)\\ 
 &&&&&&& \\[-0.1cm]
\hline
\hline
\end{tabular*}}
\caption{Quality assessment
of the semi-leptonic $K\to\pi$
vector form factor at vanishing momentum transfer, $\fplus$.
FLAG classifies the publication status as 
\gA\, published or plain update of published results,
\oP\, preprint or
\rC\, conference contribution.
\label{tab:scatter}}
\end{table}
The results for $f_K/f_\pi$ and for $\fplus$ for 
$N_f=2$, $N_f=2+1$ and $N_f=2+1+1$, respectively, are all mutually
compatible. The observation that the simulation and analysis techniques that lead to
all these results differ
significantly amongst the quoted collaborations causes confidence in the
approach. The effect of adding the dynamical strange quark (and the charm quark)
does not lead to any visible effects beyond the current level of precision.\\
The following averages/recommended values are explained in detail in the FLAG document:
$f_K/f_\pi|_{N_f=2+1}=1.193(5)$ \,(average over BMW, MILC and  HPQCD/UKQCD),\linebreak
$f_K/f_\pi|_{N_f=2  }=1.210(18)$ (ETM)  and,
$\fplus|_{N_f=2+1}   =0.959(5)$ (RBC+UKQCD) and\linebreak
$\fplus|_{N_f=2  }   =0.956(8)$ (ETM).\\
On the one hand 
these results can be used for making predictions for $|V_{ud}|$, 
$|V_{us}|$, $\fplus$ and $f_K/f_\pi$ based on the experimental results 
(\ref{eqn:masterfKpi}) and (\ref{eqn:masterKtopi})
and on the assumption of 
CKM first row unitarity $|V_{ud}|^2+|V_{us}|^2+|V_{ub}|^2=1$ (at the current
level of precision $|V_{ub}|$ is too small to play any significant role). 
On the other hand, when
using only the experimental result as input the first row unitarity can be tested.
This is summarised in the plot in figure \ref{fig:VudVus}.\\
\begin{figure}
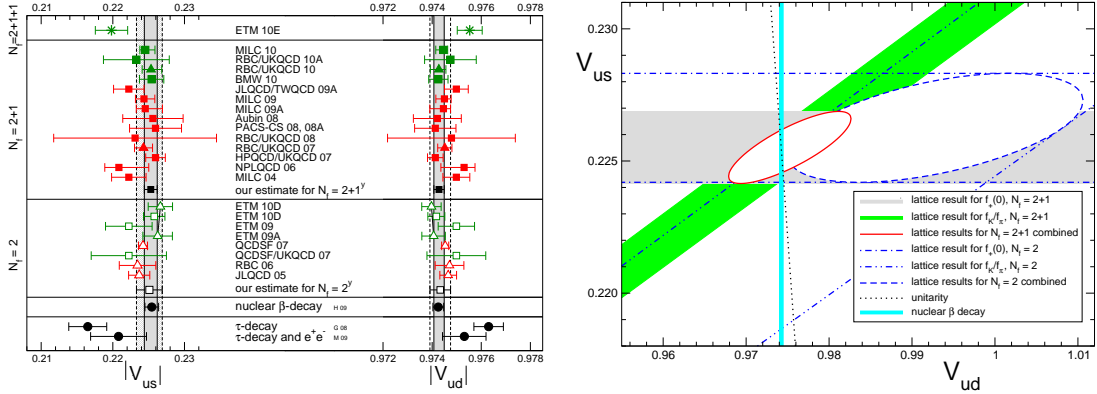

	\small 
	\begin{center}\hspace{-3mm}
	\begin{minipage}{.43\linewidth}
	{\epsfig{scale=.36,file=bildla/Vusud.eps}}
	\end{minipage}\hspace{9mm}
	\begin{minipage}{.43\linewidth}
	{\epsfig{scale=.36,file=bildla/VusVersusVud.eps}}
	\end{minipage}
	\end{center}\mbox{}\\[-8mm]
\caption{
Left: Lattice predictions for $|V_{us}|$ and $|V_{ud}|$ assuming SM unitarity.
Right: FLAG's illustration of lattice results in the $|V_{us}|$-$|V_{ud}|$-plane
\protect\cite{Colangelo:2010et}.
The ellipse represent the combined unitarity analysis for $N_f=2+1$ flavours
(solid red) and $N_f=2$ flavours (dashed blue) while the black dashed
line represents SM-unitarity. According to this analysis all results
are compatible with first row unitarity.\label{fig:VudVus}}
\end{figure}
\nocite{RBCUKQCD10, RBCUKQCD07, Lubicz:2010bv, Lubicz:2009ht, Brommel:2007wn, Dawson:2006qc, Tsutsui:2005cj}
\nocite{Farchioni:2010tb, Bazavov:2010hj, Aoki:2010dy, BMW10, JLQCD:2009sk, Bazavov:2009fk, Bazavov:2009bb, Aubin:2008ie, Aoki:2008sm, Kuramashi:2008tb, Allton:2008pn, Follana:2007uv, Beane:2006kx, Aubin:2004fs, Lubicz:2010bv, ETM09, QCDSFUKQCD}
\nocite{Leutwyler:1984je, Post:2001si,Bijnens:2003uy,Jamin:2004re, Cirigliano:2005xn, Kastner:2008ch}
Given that KLOE-2 is aiming at reducing the uncertainty in their experimental determination
for $|V_{us}|\fplus$ by a factor of about two in the near future 
\cite{Babusci:2010ym,Erica} it is fair to ask about prospects 
 on the theory side. 
Recent progress for $\fplus$ \cite{Boyle:2007wg,RBCUKQCD07,RBCUKQCD10} has allowed to 
remove one of the two most dominant uncertainties (momentum resolution
in lattice simulations). The remaining dominant uncertainty is the
one due to the chiral extrapolation which will disappear once results appear for
physical pion masses.
Cut-off effects in this observable will remain a sub-dominant uncertainty 
for a while: flavour symmetry implies that if the average 
light quark mass $m_q$ is set equal to the strange quark mass $m_s$, 
the lattice data yield $\fplus=1$, irrespective of the lattice spacing or the size of the box and for any value of $m_s$. Cut-off effects can therefore only affect the difference $1-\fplus$, which turns out to be about 0.04. 
For $f_K/f_\pi$ the error
due to the chiral extrapolation will also disappear once all collaborations
simulate directly at the physical point. The statistical error can be 
reduced by simulating longer (naively it reduces with $1/\sqrt{N}$ where $N$ 
is proportional to the Monte Carlo time). 
\subsection{Neutral Kaon mixing}\mbox{}\\[-7mm]
The mixing of neutral pseudoscalar mesons plays an important role in the 
understanding of
the physics of CP-violation. Here we present the summary of lattice
data for neutral kaon mixing which provides a probe for indirect
CP-violation.
The plot in figure \ref{fig:bK_hat} shows the status of lattice computations for
$\hat B_K$. There has been tremendous progress over the
last, say, five years. In particular the utilisation of chirally symmetric
lattice fermion formulations~\cite{Aubin:2009jh,Aoki:2010pe} has allowed to circumvent the problem of 
operator mixing of the 4-fermion operator. This reduces systematic effects considerably. 
\begin{figure}
\small 
	\begin{center}\hspace{-3mm}
	{\epsfig{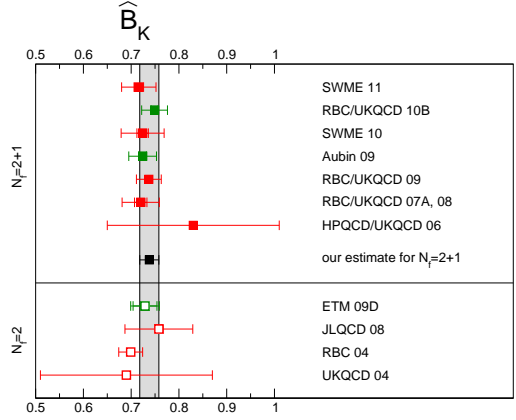}}
	\end{center}
\caption{Scatter plot for $N_f=2$ and $N_f=2+1$ results for the neutral kaon 
	mixing parameter $\hat B_K$.
\label{fig:bK_hat}}
\end{figure}
\nocite{Kim:2011qg,Aoki:2010pe,Bae:2010ki,Aubin:2009jh,Kelly:2009fp,Antonio:2007pb,Allton:2008pn,Gamiz:2006sq,Bertone:2009bu,Aoki:2008ss,Aoki:2004ht,Flynn:2004au,Albertus:2010nm,Bernard:2009wr,Blossier:2009gd,Aoki:2003xb,Bernard:2002pc,AliKhan:2001jg,Dalgic:2006gp,Simone:2010zz}
The FLAG recommended values for the renormalisation group independent $B$-parameter 
which are indicated by the vertical bands in the plot in figure~\ref{fig:bK_hat} are
 $\hat B_K=0.738(20)$ ($N_f=2+1$, Aubin 09, RBC/UKQCD 10B) and 
 $\hat B_K=0.792(25)(17)$ ($N_f=2$, ETM 09D). 
\subsection{Quark masses}\mbox{}\\[-7mm]
The bare parameters of the $N_f=2(2+1)$ Lagrangian are the gauge coupling
and the up- and down- (and strange) quark masses. Since lattice 
QCD is simulated in the iso-spin limit, it is sufficient to provide
two (three) hadronic quantities from experiment to tune the parameters
of the Lagrangian to their physical values. There exist well defined
procedures for how to determine renormalised quark masses in a second step.
Various collaborations have carried out programmes to determine
the light quark masses and the scatter plot in figure~\ref{fig:quarkmasses}
gives an impression for the precision which 
can by now be achieved in lattice QCD. The FLAVIA recommended values indicated by
the vertical bands in the plots are 
$m_{ud}=3.43(11)$MeV, $m_s=94(3)$MeV for $N_f=2+1$ (MILC 09A, RBC/UKQCD 10A and HPQCD 10) and
$m_{ud}=3.6(2)$MeV, $m_s=95(6)$MeV for $N_f=2$ (ETM 10B).
\begin{figure}
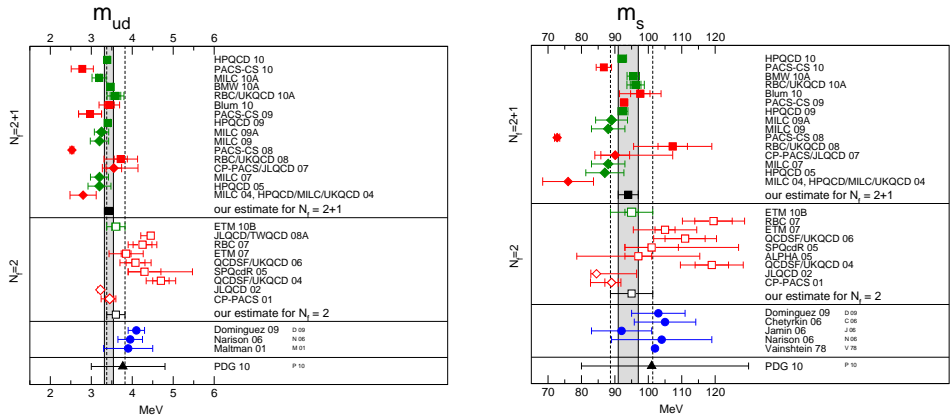

\small 
	\begin{center}\hspace{-3mm}
	\begin{minipage}{.43\linewidth}
	{\epsfig{scale=.35,file=bildla/mud.eps}}
	\end{minipage}
	\begin{minipage}{.43\linewidth}
	{\epsfig{scale=.35,file=bildla/ms.eps}}
	\end{minipage}
	\end{center}
\caption{Scatter plot for the average up- and down-quark masses and the strange
quark mass.
\label{fig:quarkmasses}}
\end{figure}
\nocite{Blossier:2010cr,
Noaki:2008iy,
Blum:2007cy,
Blossier:2007vv,
Gockeler:2006jt,
Becirevic:2005ta,
DellaMorte:2005kg,
Gockeler:2004rp,
Aoki:2002uc,
AliKhan:2001tx,
Aoki:2010wm,
Bazavov:2010yq,
McNeile:2010ji,
Durr:2010vn,Durr:2010aw,
Aoki:2010dy,
Blum:2010ym,
Aoki:2009ix,
Davies:2009ih,
Bazavov:2009fk,
Bazavov:2009bb,
Aoki:2008sm,
Allton:2008pn,
Ishikawa:2007nn,
Mason:2005bj,
Aubin:2004fs,Aubin:2004ck,Nakamura:2010zzi,Narison:2005ny,Maltman:2001nx,Dominguez:2008jz}
In fact, the precision is so good
 that QCD and QED iso-spin breaking effects play
a significant role and need to be taken into account in any reliable
estimate of the systematic uncertainties. The combined statistical 
and systematic errors in the FLAG average for $N_f=2+1$ 
are around 3\% for both the average
up- and down-quark mass and the strange quark mass. 
FLAG also provides estimates of the individual 
up- and down-quark masses and found, based on phenomenological estimates of the
quark's electro-magnetic self-energies that the up-quark mass to be different from
zero by 15 standard deviations.
\subsection{Low energy constants}\mbox{}\\[-7mm]
On the one hand lattice practitioners are putting a lot of effort into improving
algorithms and computers in order to be able to simulate QCD for physical quark
masses, in this way avoiding systematic-affected extrapolations to the 
physical point. That this is now possible has been demonstrated only
recently \cite{Durr:2010aw}. However, many simulations are carried out with
heavier than physical quark masses and chiral perturbation theory is employed to 
provide ans\"atze for the mass dependence of the observable under consideration.
The low-energy constants of chiral perturbation theory can be determined from
fits of these ans\"atze to the lattice data. The FLAG document also presents 
an overview and analysis over the status of determinations of these low-energy 
constants.\\[-10mm]
\section{Outlook}\mbox{}\\[-8mm]
FLAG has set out to provide the (Beyond) Standard Model phenomenologist
with predictions for low-energy parameters and observables of QCD. The aim
is to screen and judge all available lattice data from a 
specialist's point of view and to provide a summary of the status and a quality 
assessment of each single result for  the non-specialist and where 
appropriate also an average or recommended range. 

FLAG is envisaging regular updates of the summary table and scatter plots
and also of the averages. In addition FLAG is working on extending 
the set of data considered so far towards observables and parameters 
related to charm and bottom quarks.\\[2mm]
\acknowledgements{\mbox{}\\[-15mm]
Many thanks to all involved in organising this interesting conference and 
for inviting FLAG. 
The speaker would also like to thank the prize-committee for choosing his talk!
}\\[-20mm]
\bibliographystyle{aipproc}
\bibliography{hadron2011}
%
%
}  
